# Molecular-Orbital Degeneracy Lifting and Suppression of a Flat-Band Metal in a Tetrahedral Cluster System NbSeI

Keita Kojima[1,2]*, Youichi Yamakawa[3], Ryutaro Okuma[1], Shunsuke Kitou[4], Taishun Manjo[5], Hayato Takano[1], Jun-ichi Yamaura[1], Yusuke Tokunaga[4], Taka-hisa Arima[4], and Yoshihiko Okamoto[1]*

[1] Institute for Solid State Physics, the University of Tokyo, Kashiwa 277-8581, Japan

[2] Graduate School of Environmental, Life, Natural Science and Technology, Okayama University, Okayama 700-8530, Japan

[3] Department of Physics, Nagoya University, Nagoya 464-8602, Japan

[4] Department of Advanced Materials Science, the University of Tokyo, Kashiwa 277-8561, Japan

[5] Japan Synchrotron Radiation Research Institute (JASRI), SPring-8, Hyogo 679-5198, Japan

* Corresponding author: keita-kojima@okayama-u.ac.jp
yokamoto@issp.u-tokyo.ac.jp

## Abstract

The lifting of degenerate electronic states has been extensively studied in transition metal compounds, where diverse quantum phenomena arise from the degrees of freedom of *d* electrons on individual atoms. In contrast, transition metal compounds containing high-symmetry clusters composed of multiple transition-metal atoms are expected to host a broader range of emergent phenomena due to the entanglement of the electronic degrees of freedom across multiple atoms. Here, we report the discovery of two distinct structural responses associated with molecular-orbital-degeneracy lifting in NbSeI, which comprises $Nb_4$ tetrahedral clusters with molecular orbital degrees of freedom and whose cubic average crystal structure is predicted to host a flat-band metallic state. Below 106 K, NbSeI is a molecular-orbital-ordered nonmagnetic insulator with orthorhombic symmetry. Above this temperature, the average structure becomes face-centered cubic without superlattice formation, while local symmetry breaking associated with pronounced local distortions of the $Nb_4$ tetrahedra lifts the molecular-orbital degeneracy. This local symmetry breaking occurs without cooperative long-range order and stabilizes a nonmagnetic insulating state instead of the flat-band metallic state predicted for the cubic average structure.

## Introduction

Orbital degrees of freedom of *d* electrons play an important role in a wide range of electronic phenomena in transition-metal compounds[1–13]. When degenerate *d* orbitals of a transition-metal atom located at a high-symmetry position are partially occupied, their degeneracy is often lifted by the cooperative Jahn-Teller effect, giving rise to orbital ordering and associated structural and electronic transitions. Various orbital-ordered states have been established, including ferroic and antiferroic orders[2,9–11] and those accompanied by multimer formation[8,12,13]. In a few materials, alternative mechanisms for lifting orbital degeneracy without conventional orbital order have also been proposed[14-17], although subsequent studies have substantially revised some of these interpretations[18–23]. The lifting of atomic orbital degeneracy and the resulting electronic states therefore constitute a central theme of conventional orbital physics based on the atomic *d* orbitals of individual transition-metal ions.

On the other hand, degenerate molecular orbitals can also emerge in highly symmetric clusters composed of multiple transition-metal atoms. In Chevrel-phase compounds containing octahedral $Mo_6$ clusters, for example, their electronic structures have long been understood in terms of cluster molecular orbitals, including degenerate molecular-orbital states[24]. In lacunar spinels $AM_4X_8$ containing tetrahedral $M_4$ clusters, partially occupied molecular orbitals give rise to diverse electronic phenomena, including cluster Jahn–Teller transitions, unconventional magnetism with topologically nontrivial spin textures, and orbital and charge redistribution within the clusters[1,25–29]. Related molecular-orbital physics is also found in compounds containing triangular $M_3$ clusters. In the $Nb_3X_8$ family, particularly $Nb_3Cl_8$, preformed $Nb_3$ trimers form the relevant low-energy electronic units, giving rise to cluster-Mott states and nearly flat electronic bands[30,31]. In $LiZn_2Mo_3O_8$, each $Mo_3$ molecular unit carries a localized spin, giving rise to frustrated magnetism[32,33]. Thus, molecular-orbital degrees of freedom in transition-metal clusters are increasingly recognized as a platform for diverse electronic phenomena. However, how molecular-orbital degeneracy is lifted in cluster-based systems and what electronic states emerge from this process remain much less explored than in conventional atomic-orbital systems.

In this paper, we report that NbSeI exhibits two distinct structural responses associated with molecular-orbital degeneracy lifting: molecular-orbital ordering and local symmetry breaking of $Nb_4$ tetrahedra without cooperative long-range order, both of which suppress the average-structure flat-band metallic state and stabilize nonmagnetic insulating states. NbSeI has been reported to crystallize in the MoSBr-type structure with cubic $F\bar{4}3m$ symmetry[34,35]. As shown in Fig. 1(a), this crystal structure can be regarded as an A-site-deficient and mixed-anion spinel, where the Nb atoms are coordinated by $Se_3I_3$ octahedra. Each Nb, Se, and I sublattice forms a breathing pyrochlore structure with alternating large and small regular tetrahedra[36]. One of the characteristics feature is the strong breathing; the ratio of the side length of the large and small $Nb_4$ tetrahedra, $d'/d$, reaches 1.6[29], which is considerably larger than those of lacunar spinels ($d'/d$ ~ 1.4)[37,38] and A-site-ordered spinels ($d'/d$ ~ 1.1)[36,39].

The pronounced breathing ratio suggests that the electronic structure of NbSeI is governed by molecular orbitals formed within the small $Nb_4$ tetrahedra rather than by atomic orbitals on individual Nb atoms, as illustrated in Fig. 1(b)[1,25]. In the local $C_{3v}$ ligand field of the $NbSe_3I_3$ octahedron, the Nb 4*d* $t_{2g}$ orbitals split into nondegenerate $a_1$ and doubly-degenerate *e* atomic orbitals. The substantially shorter Nb–Nb distances within the small tetrahedra promote strong hybridization of these Nb 4*d* orbitals, giving rise to well-defined molecular orbitals localized on each $Nb_4$ tetrahedron. Since each Nb atom contributes two 4*d* electrons, the resulting molecular orbitals accommodate eight 4*d* electrons per $Nb_4$ tetrahedron, leaving two electrons in the triply-degenerate antibonding $t_2$ molecular orbitals, as shown in Fig. 1(b). These partially occupied $t_2$ molecular orbitals are of particular importance because they provide the relevant orbital degrees of freedom. Notably, these molecular-orbital degrees of freedom are rooted in the antibonding combination of the four $a_1$ atomic orbitals of the $Nb_4$ tetrahedron, rather than in the atomic-orbital degeneracy of an isolated Nb ion. Despite these distinctive molecular-orbital characteristics, previous studies of NbSeI have been limited to its room-temperature crystal structure and magnetization, leaving its electronic and structural properties largely unexplored[34,35].

## Results

### Formation of $Nb_4$ tetrahedral clusters and an average-structure flat band

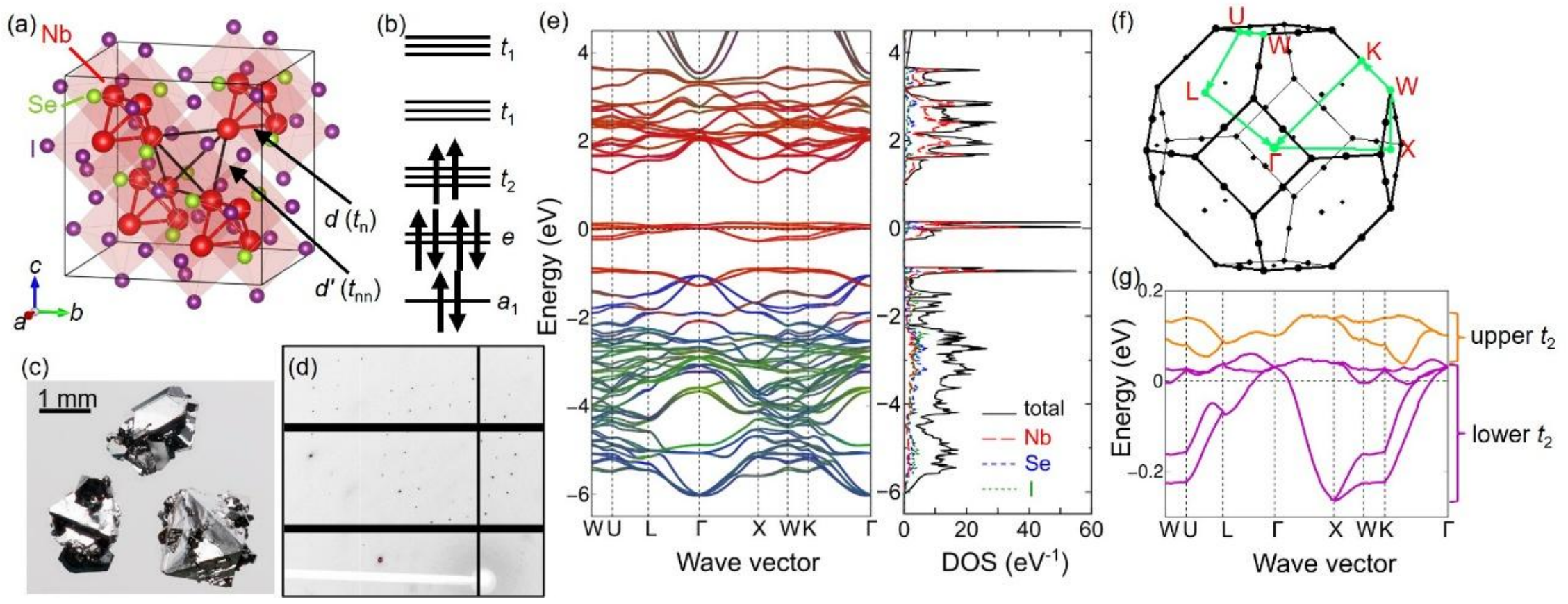

**Fig. 1. Crystal and electronic structures of NbSeI.** (a) Crystal structure. (b) Energy diagram of the molecular orbitals of a $Nb_4$ tetrahedron made of $t_{2g}$ orbitals of Nb 4*d* orbitals. (c) NbSeI single crystals. (d) A single crystal XRD pattern measured at 300 K. (e) Electronic structure and density of states calculated with spin–orbit coupling using refined structural parameters at 300 K. The $E_F$ is set to 0 eV. The red, blue, and green colors in the electronic structures represent the contributions of Nb, Se, and I atoms, respectively. (f) The Brillouin zone of the face centered cubic lattice. The green line corresponds to the path in the band dispersion in Fig. 1(e). (g) Enlarged view of the electronic structure shown in (e) near the $E_F$. The bands were split into the upper (orange) and lower (purple) $t_2$ bands by the spin–orbit coupling. Because NbSeI lacks inversion symmetry, spin degeneracy is generally lifted by spin–orbit coupling, although symmetry-protected twofold degeneracies remain along Γ–X for all bands and along Γ–L for some bands.

We synthesized single crystals of NbSeI (see Materials and Method section for details). The obtained single crystals had an octahedral shape with edges of at most 2 mm, as shown in Fig. 1(c). Structural refinement using the single-crystal X-ray diffraction (XRD) data, as shown in Fig. 1(d), yielded a sufficiently low reliability factor of $R_{I > 3\sigma}$ = 3.29% for the cubic $F\bar{4}3m$ symmetry at 300 K. Details of the structural analysis and structural parameters are provided in the Supporting Information. Regardless of the initial conditions considering the intersite defects or vacancies of Nb, Se, and I atoms, the refinements converged in a model in which each atomic site is fully occupied by Nb, Se, or I atoms. The shorter and longer Nb–Nb bond lengths are $d$ = 2.9754(4) Å and $d'$ = 4.6463(4) Å, respectively, yielding large $d'/d$ ratio of 1.56, indicative of pronounced breathing distortion and consistent with $d'/d$ = 1.6 reported in the previous study[35]. Focusing on the Se and I atoms coordinated to Nb, the Nb–Se and Nb–I bond lengths in a $NbSe_3I_3$ octahedron are 2.5377(3) Å and 3.00804(19) Å, respectively. The Nb–Se bond length is significantly shorter than the sum of the ionic radii of $Nb^{3+}$ and $Se^{2-}$ ($Nb^{3+}$: 0.72 Å and $Se^{2-}$: 1.98 Å)[40], indicating a strong covalent character of the Nb–Se bond. In contrast, the Nb–I bond length is slightly longer than the sum of the ionic radii of $Nb^{3+}$ and $I^-$ ($I^-$: 2.20 Å)[40], suggesting a more ionic nature, as well as the chemical pressure arising from short Nb–Se bonds.

Figures 1(e) and 1(g) show the results of the first-principles calculations for NbSeI using the refined structural parameters at 300 K, with the Brillouin zone shown in Fig. 1(f). An important feature is the formation of a narrow energy band at the Fermi energy $E_F$. As shown in Fig. 1(g), this band can accommodate six electrons, reflecting the $t_2$ molecular orbitals, which is supported by the tight-binding results discussed later. This $t_2$ band is split into two bands, the upper and lower $t_2$ bands with higher and lower energies, respectively, due to the spin–orbit coupling. The lower $t_2$ band can accommodate four electrons, and two 4$d$ electrons occupy half the band. Note that this band forms a quadruply-degenerate $\Gamma_8$ state at the Γ point, as required by the $F\bar{4}3m$ symmetry, resulting in the band dispersion necessarily crossing $E_F$. These results indicate that the average cubic structure of NbSeI is predicted to host a flat-band metallic state, which we refer to as an “average-structure flat-band metal”. The implications of this result for the experimentally observed insulating state are discussed below. In particular, the dispersion at the top of the lower $t_2$ band, which lies approximately 30 meV above $E_F$, is almost perfectly flat. Such flat dispersion close to $E_F$ has never been achieved in breathing pyrochlore materials, although similar narrow $t_2$ bands have been predicted in $GaV_4S_8$ and ReSTe[25,41].

## Molecular-orbital ordering in Phase II

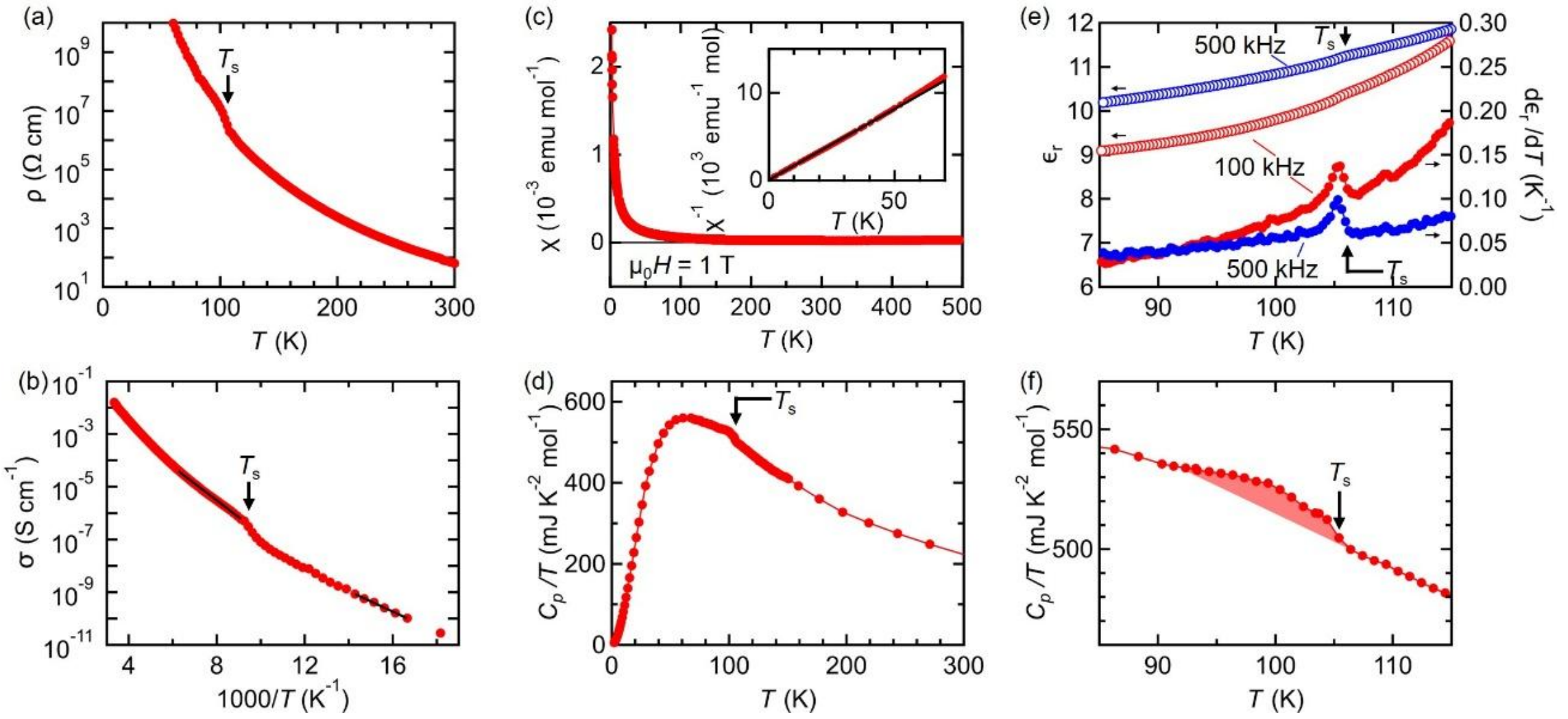

**Fig.2 Physical properties of NbSeI single crystals.** (a) Temperature dependence of electrical resistivity ρ. (b) An Arrhenius plot of electrical conductivity σ. (c) Temperature dependence of field-cooled magnetic susceptibility measured in a magnetic field of 1 T. (d) Temperature dependence of heat capacity divided by temperature. The inset shows the result of the Curie-Weiss fitting to $\chi^{-1} = (T - \theta_W)/C$ between 2 and 10 K, yielding a Curie constant of $C = 0.00578(2)$ emu K mol$^{-1}$, Weiss temperature of $\theta_W = -0.40(8)$ K, and $\chi_0 = 3(2)\times10^{-5}$ emu mol$^{-1}$. This $C$ value corresponds to an effective magnetic moment of $\mu_{eff} = 0.215$ $\mu_B$ per Nb atom. (e) Temperature dependence of the relative dielectric constant and its temperature derivative. (f) Enlarged view of the heat capacity divided by temperature.

In contrast to the average-structure flat-band metallic state predicted by first-principles calculations, the electrical resistivity ρ of a single crystal exhibits a semiconducting behavior, accompanied by a small anomaly at $T_s$ = 106 K, as shown in Fig. 2(a). The Arrhenius plots of electrical conductivity $\sigma = \rho^{-1}$ above $T_s$ (110 K < $T$ < 160 K) and below $T_s$ (60 K < $T$ < 70 K) shown in Fig. 2(b) yield activation energies of 127 and 77 meV, respectively. As shown in Fig. 2(c), the magnetic susceptibility χ, measured up to 500 K, remains small and nearly temperature independent above $T_s$ at high temperatures and exhibits only a Curie-like increase toward 0 K, corresponding to approximately 1.5% of Nb atoms carrying localized $S$ = 1/2 spin. These results indicate that NbSeI remains a nonmagnetic insulator with only a small concentration of impurity spins over the entire measured temperature range up to 500 K, with no evidence for the emergence of a metallic state.

As shown in Figs. 2(d) and 2(f), the heat capacity divided by temperature, $C_p/T$, of NbSeI

forms a broad peak below $T_s = 106$ K, indicating the presence of a phase transition at $T_s$. As shown in Fig. 2(e), an anomaly at $T_s$ also appears in the relative dielectric constant $\varepsilon_r$ data, although the small anomaly at $T_s$ indicates that the transition does not involve a change in dielectric properties. Hereinafter, the phases in the temperature ranges above and below $T_s$ are named Phases I and II, respectively.

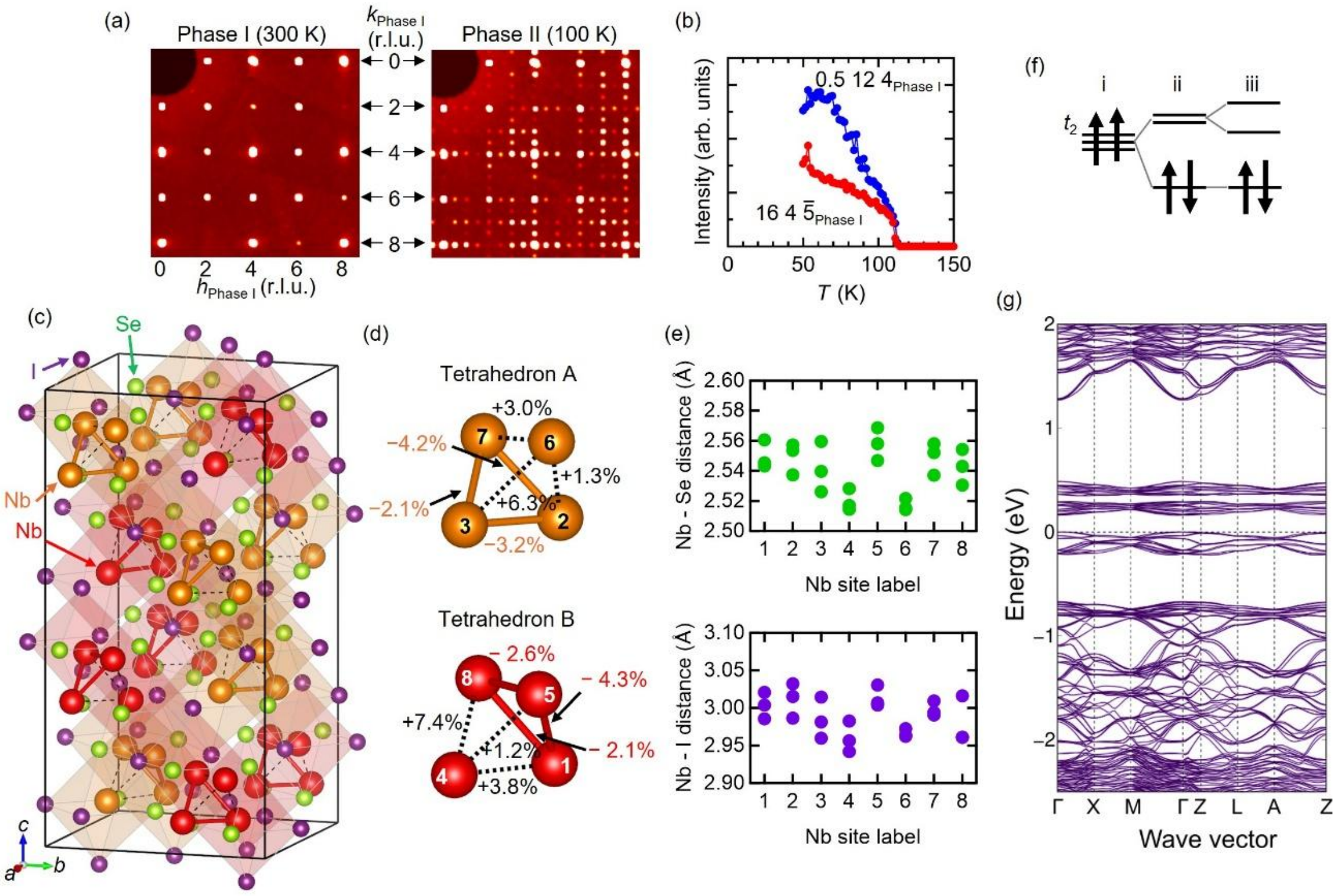

**Fig. 3. Crystal and electronic structures of Phase II of NbSeI.** (a) Single crystal XRD patterns of 300 K (Phase I) and 100 K (Phase II) at the *hk*0 plane. (b) Temperature dependences of the intensities of (0.5 12 4) and (16 4 $\bar{5}$) superlattice reflections. (c) Unit cell of Phase II. Four Nb sites forming tetrahedra A are shown in orange, whereas the other four Nb sites forming tetrahedra B are shown in red. (d) Distorted $Nb_4$ tetrahedra A (upper) and B (lower) at 100 K (Phase II). The relative length changes from the average structure of Phase I for each edge is shown. (e) Nb-Se and Nb-I interatomic distances at 100 K. Each site label is defined in Fig. 3(d) and Supplementary Table 4 in Supporting Information. (f) Energy diagram of $t_2$ orbitals in cubic symmetry (i), that with the 3-in-1-out distortion (ii), and that with the 3-in-1-out distortion and additional orthorhombic distortion (iii). (g) Electronic structure of Phase II calculated with spin–orbit coupling. The $E_F$ is set to 0 eV. The three bands in the energy range between –0.2 and 0.5 eV correspond to the energy levels shown in (f), iii.

As shown in Fig. 3(a), superlattice peaks are observed at 100 K. When the diffraction peaks of Phase II are indexed using the same unit cell as Phase I, reflections with half-integer indices, such as (4 0.5 0), appear, indicating that the unit cell is double along at least one crystallographic direction. However, no reflections corresponding to doubling along two or three directions, such as (4.5 0.5 0) or (4.5 0.5 0.5), are observed. These results suggest that the unit cell of Phase II is not a 2×2×1 or 2×2×2 expansion of Phase I, but rather a 2×1×1 superstructure.

In addition, reflections violating the extinction conditions of a face-centered lattice are observed below $T_s$. Assuming a 2×1×1 unit cell with multiple domains, systematic absences are observed for $h00$, $0k0$, and $00l$ reflections with $h$, $k$ and $l$ all odd. On the other hand, no splitting of the fundamental reflections is detected in Phase II. Notably, even when cubic symmetry is assumed, the agreement among the intensities of symmetry-equivalent reflections remains reasonably good, with a deviation of only 3.8%. Therefore, the crystal system cannot be uniquely determined at this stage.

Based on Table 3.1.4.1 in *International Tables for Crystallography, Volume A*[42], we identified candidate space groups that are subgroups of the parent $F\bar{4}3m$ symmetry and satisfy the observed extinction conditions. The possible space groups include the tetragonal $P\bar{4}2_1m$, $P\bar{4}2m$, and $P\bar{4}$, as well as the orthorhombic $P2_12_12_1$, $P2_12_12$, $P222_1$, and $P222$. Structural refinements were performed for eighteen models differing in the origin choice and equivalent orthorhombic settings (e.g., $P222_1$ vs $P2_122$). Among these, only the model based on the orthorhombic space group $P2_12_12_1$, shown in Fig. 3(c), yielded sufficiently low $R$ values of $R_{I>3\sigma} = 4.62\%$. We therefore conclude that the space group of Phase II is orthorhombic $P2_12_12_1$ with a doubling of the unit cell of Phase I. As a result of the unit-cell doubling and symmetry lowering from Phase I, each element occupies eight crystallographically distinct sites in Phase II. Detailed structural parameters are provided in the Supporting Information.

To examine whether the site splitting is accompanied by charge disproportionation, we analyzed the cation–anion bond distances within each $NbSe_3I_3$ octahedron. The Nb-site dependent Nb–Se and Nb–I bond distances are plotted in Fig. 3(e), where the site labels correspond to those defined in Fig. 3(d) and Supplementary Table 4 in Supporting Information. As shown in Fig. 3(e), most Nb–Se bond lengths are nearly identical across the different Nb sites. The Nb–Se bond lengths for the Nb4 and Nb6 sites show slight deviations from the others; however, these differences are not pronounced. These differences most likely originate from structural distortions rather than charge redistribution, indicating that no charge disproportionation occurs in Phase II.

Despite the absence of charge disproportionation and the negligibly small orthorhombic distortion in Phase II, the $Nb_4$ tetrahedra are heavily distorted. As discussed earlier, eight Nb sites exist in Phase II, forming two inequivalent $Nb_4$ tetrahedra A and B. As shown in Fig. 3(d), the distortions of both tetrahedra exhibit a 3-in-1-out type distortion, consisting of three shorter edges forming a triangle (thick solid lines) and three other longer edges (broken lines). The shorter and longer edges are 2–5% shorter and 1–8% longer than those in Phase I, respectively. The presence of this considerable distortion strongly suggests that cooperative Jahn–Teller distortion occurs in Phase II. Although this 3-in-1-out geometry resembles the trimerized structures found in β-pyrochlore $CsW_2O_6$[43] and lacunar spinel $GaNb_4Se_8$[38], both of which are accompanied by charge disproportionation within the tetrahedron. Trimerized structure also seen, as well as those of the triangular lattice compounds $LiVO_2$ and $LiVS_2$[44-48], the present distortion should not be interpreted as the formation of an independent $Nb_3$ trimer. Instead, it reflects molecular-orbital ordering driven by the lifting of the degeneracy of the pre-existing $Nb_4$ molecular orbitals.

To understand how the observed distortion lifts the molecular-orbital degeneracy, we first consider an ideal 3-in-1-out configuration. In this case, triply-degenerate $t_2$ molecular orbitals [Fig. 3(f), i] are split into two high-energy and one low-energy molecular orbitals [Fig. 3(f), ii], as discussed below using the tight-binding model. However, as shown in Fig. 3(g), the first-principles calculations reveal an additional splitting of the high-energy orbitals, reflecting a further symmetry lowering of each tetrahedron from the ideal 3-in-1-out configuration. This splitting is schematically illustrated in Fig. 3(f), iii. The two 4*d* electrons per $Nb_4$ tetrahedron doubly occupy the low-energy molecular orbital, resulting in a nonmagnetic insulator. Thus, the electronic state of Phase II can be understood as a molecular-orbital-ordered nonmagnetic insulator.

## Local Jahn–Teller-like symmetry breaking in Phase I

The phase transition at $T_s$ differs markedly from a conventional cooperative Jahn–Teller transition between orbital-degenerate and orbital-ordered states. The entropy change $\Delta S$ associated with the phase transition is estimated to be ~ 0.1 J $K^{-1}$ $mol^{-1}$ from Fig. 2(f), which is much smaller than those of typical Jahn–Teller transitions as well as orbital entropy for the triply degenerate $t_2$ orbitals [$(k_B N_A \ln 3)/4$ = 2.3 J $K^{-1}$ $mol^{-1}$, where $k_B$ and $N_A$ are Boltzmann and Avogadro constants, respectively] [49,50]. The ρ and χ measurements indicate that both Phases I and II are nonmagnetic insulating phases. These results suggest that the degeneracy of $t_2$ molecular orbitals is already lifted in Phase I.

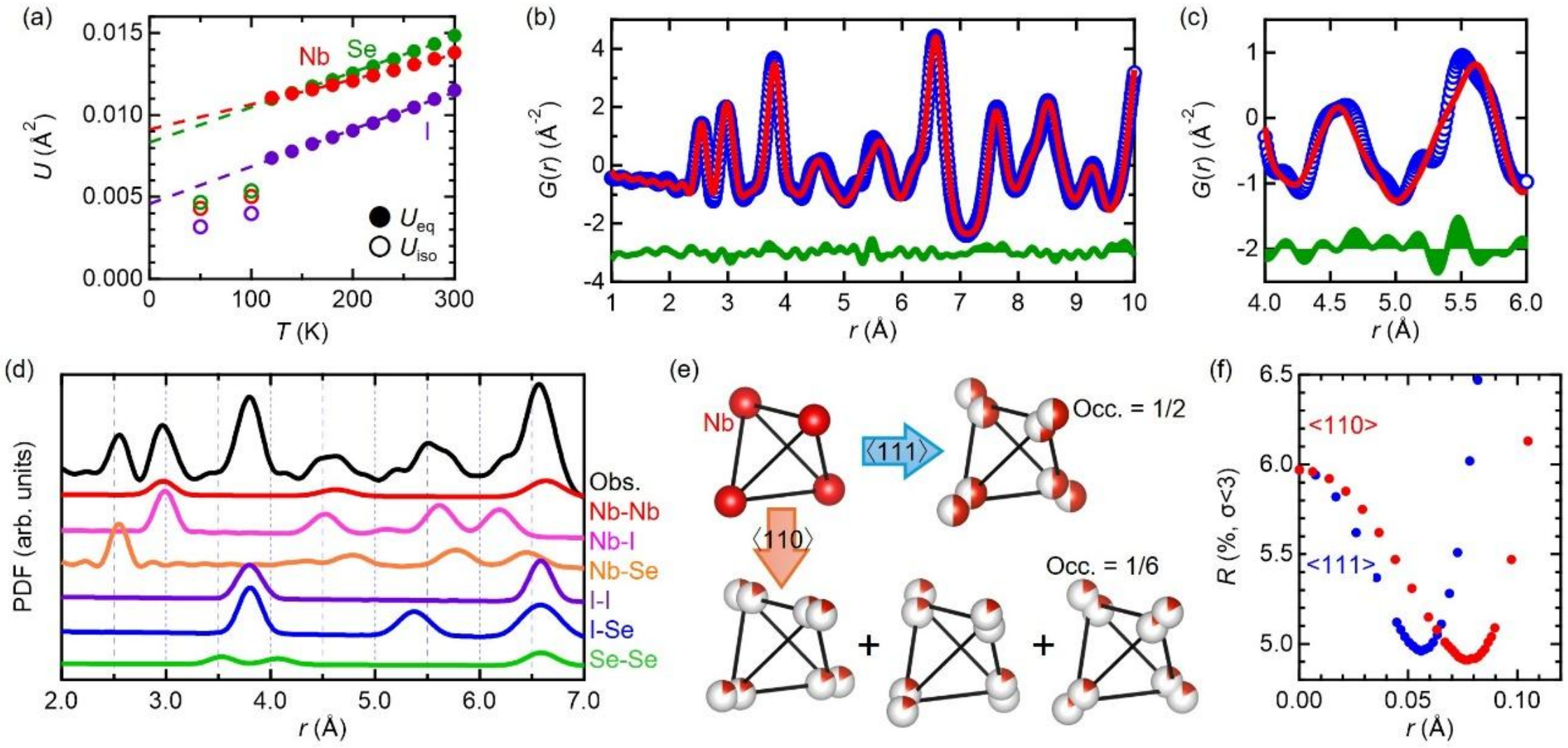

**Fig. 4. Local distortion in Phase I of NbSeI.** (a) Temperature dependence of the thermal displacement parameter for each atomic site. Phase I data are plotted using $U_{eq}$, whereas Phase II data are plotted using $U_{iso}$ due to the restriction of the structural analysis. (b,c) Results of the pair distribution function (PDF) analysis at 300 K assuming the cubic average structure. Blue circles, red lines, and green shaded curves represent the observed PDF, the PDF calculated using the cubic average structure, and the corresponding residuals, respectively. The weighted reliability factor is $R_w$ = 11.6%. (d) Observed PDF at 300 K and simulated partial PDF contributions from the individual atomic pairs. (e) Split sites for the Nb sites along the ⟨111⟩ and ⟨110⟩. (f) Reliability factor $R_{I>3\sigma}$ in the structural refinement according to the split-site model for the 160 K data, as a function of displacement from the regular position.

Detailed analyses of the XRD data in Phase I provide evidence for local structural distortions. As shown in Fig. 4(a), the atomic displacement parameters ($U_{eq}$) of all atomic sites do not extrapolate to zero at 0 K. Because no crystallographic defects were detected at any atomic site, this behavior indicates the presence of local distortion.

To further investigate the local structure in Phase I, we performed one-dimensional pair distribution function (PDF) analysis at 300 K. Figure 4(b) shows the fit of the observed PDF using the cubic average structure. Although the overall PDF profile is reasonably reproduced by the average structure, a closer inspection around $r \sim 5.5$ Å [Fig. 4(c)] reveals a clear mismatch between the observed and calculated profiles, providing direct evidence that the cubic average structure does not fully describe the local interatomic distances.

To identify the origin of this discrepancy, we calculated the partial PDF contributions from the individual atomic pairs [Fig. 4(d)]. The peak around $r \sim 5.5$ Å consists of overlapping contributions from the Nb–I, Nb–Se, and Se–I atomic pairs. Similar overlaps are also found for other characteristic PDF peaks. Therefore, the present PDF analysis alone does not allow the atomic pair primarily responsible for the deviation from the average structure to be identified.

We therefore performed a split-site model analysis to further evaluate the local atomic displacements (see Supporting Information for details). The analysis was carried out for the Phase I at 160 K by introducing split Nb sites while preserving the center of mass of each crystallographic site [Fig. 4(e)]. As shown in Fig. 4(f), the reliability factor, $R$, is significantly reduced when the Nb sites are split along either the $\langle 111 \rangle$ or $\langle 110 \rangle$, reaching the minimum values for displacements of $r$ = 0.056 Å and 0.077 Å, corresponding to ±1.9% and ±2.6% of the $Nb_4$ tetrahedral edges length, respectively. These displacement amplitudes are comparable to the magnitude of the Jahn–Teller distortion observed in Phase II. Applying the same analysis to the Se site also improves the refinement, whereas the same treatment of the I sites increases the $R$ value (see Supporting Information). Because the Se atoms bridge the Nb–Nb edges whereas the I atoms are located outside the $Nb_4$ tetrahedra, as shown in Fig. 1(a), this result provides direct evidence that the local distortion is associated with distortions of the $Nb_4$ tetrahedra. The displacement amplitudes obtained from the split-site model are also consistent with those estimated from the atomic displacement parameters [$\sqrt{U} \sim 0.1$ Å, as shown in Fig. 4(a)]. Taken together, these results indicate the presence of substantial local Nb displacements in Phase I. Their magnitude is comparable to that observed in Phase II, supporting the presence of locally distorted $Nb_4$ tetrahedra while preserving the average cubic symmetry.

Weak diffuse scattering with an intensity less than $10^{-4}$ of the corresponding Bragg reflections is observed around the Bragg peaks in the XRD patterns (Supplementary Fig. 1). The extremely weak diffuse scattering rules out the development of pronounced short-range order associated with the local distortions, but does not allow us to determine whether short-range order with a very short correlation length is present. Likewise, the detailed nature of the local distortion, including the displacement direction and the relationship between the distortions in Phases I and II, remains unclear. Moreover, the present diffraction measurements do not distinguish between static and dynamic local distortions.

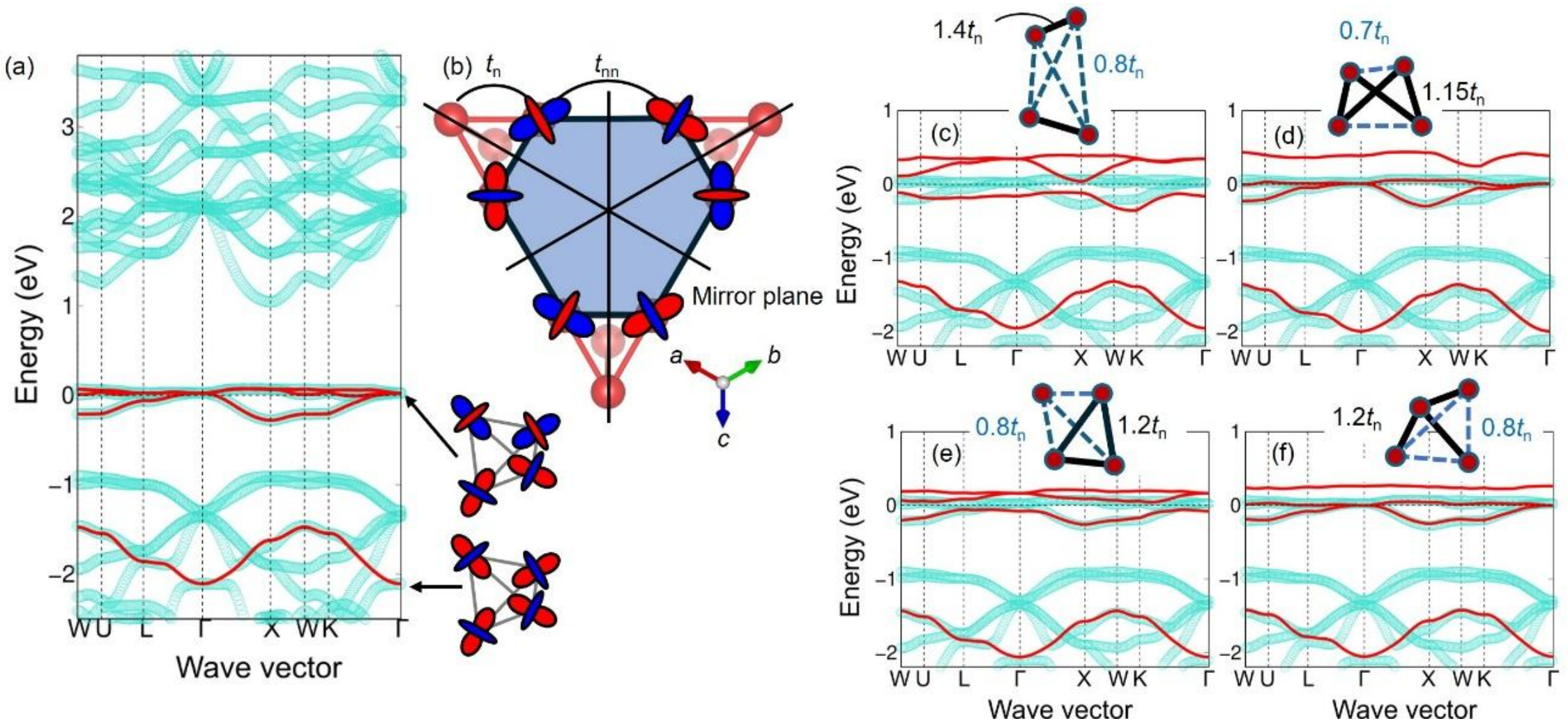

**Fig. 5. Electronic structure based on the tight-binding model using four $a_1$ atomic orbitals in a $Nb_4$ tetrahedron (red curves).** The cyan curves are the first-principles calculation result without spin–orbit coupling using refined structural parameters at 300 K. (a) Tight-binding model result. The nearest neighbor and next nearest neighbor hopping $t_n$ = –0.42 eV and $t_{nn}$ = –0.05 eV, respectively. Energy dispersions at approximately $E_F$ and –2 eV correspond to antibonding and bonding molecular orbitals made of the $a_1$ atomic orbitals, respectively. (b) Compact localized state formed on a $Nb_6$ hexagon. (c-f) Effects of the distortion of $Nb_4$ tetrahedra on the tight-binding results. (c) and (d) correspond to the uniaxial elongation and compression along *a*, *b*, or *c* directions, respectively. (e) and (f) correspond to the 3-in-1-out-type and 1-in-3-out-type distortions, respectively.

The electronic structure of the average structure of Phase I is well described by the combined first-principles and tight-binding calculations. As shown in Fig. 5(a), the $t_2$ band at $E_F$ obtained from the first-principles calculations is accurately reproduced by a tight-binding model consisting of one Nb 4*d* orbital (the $a_1$ atomic orbital under the local $C_{3v}$ ligand field) on each Nb atom. The nearest-neighbor hopping within the $Nb_4$ tetrahedron and the next-nearest-neighbor hopping between neighboring tetrahedra are estimated to be $t_n$ = –0.42 eV and $t_{nn}$ = –0.05 eV, respectively. These results demonstrate that the $t_2$ band is well described by antibonding molecular orbitals constructed from the four $a_1$ orbitals of the $Nb_4$ tetrahedron and that the inter-tetrahedron coupling is exceptionally weak ($t_{nn}/t_n$ = 0.1). This reflects the strongly breathing pyrochlore structure ($d'/d$ = 1.56) and places NbSeI close to the nearly isolated tetrahedral-cluster limit, providing an ideal platform for investigating molecular-orbital physics.

The nearly flat dispersion near the top of the lower $t_2$ band, as shown in Fig. 1(g), originates primarily from the quantum-mechanical destructive interference inherent to the pyrochlore geometry, as in other pyrochlore-based flat-band systems[51-53]. As illustrated in Fig. 5(b), the $a_1$ atomic orbitals form a compact localized state on a hexagonal loop of the breathing pyrochlore structure. Because every Nb site connected to this hexagon lies on a mirror plane passing through the loop, the hopping amplitudes from the compact localized state cancel exactly by destructive interference, giving rise to an ideally flat band. Although destructive interference gives rise to a flat band, further-neighbor hopping generally introduces a finite dispersion. In NbSeI, however, the exceptionally weak inter-tetrahedron coupling substantially suppresses such further-neighbor hopping. This additional suppression preserves the nearly ideal flatness of the lower $t_2$ band obtained in the first-principles calculations.

Figures 5(c)–5(f) show the tight-binding results on the changes in the electronic state when the $Nb_4$ tetrahedra are distorted. Among these four distortion modes, nonmagnetic insulating states can be realized in the cases of Figs. 5(c) and 5(e). The distortion mode in Fig. 5(e) corresponds to Phase II, where the $Nb_4$ tetrahedra are distorted into a 3-in-1-out configuration (as discussed above, the actual Phase II exhibits further symmetry reduction). At present, the exact nature of the distortion in Phase I is not yet identified. However, these results demonstrate that sufficiently large local symmetry breaking of the $Nb_4$ tetrahedra lifts the molecular-orbital degeneracy and stabilizes a nonmagnetic insulating state, even in the absence of long-range order. If the molecular-orbital degeneracy is already lifted by local symmetry breaking in Phase I, only a small entropy change is expected at the transition at $T_s$, consistent with the experimental observation.

**Discussion**

Phase I is a nonmagnetic insulating phase despite the absence of the cooperative long-range molecular-orbital order observed in Phase II. The structural analyses provide multiple lines of evidence for local symmetry breaking of the $Nb_4$ tetrahedra in Phase I: the Nb sites exhibit large $U_{eq}$, the cubic average structure does not fully reproduce the local PDF, and split-site refinements on the Nb and Se sites significantly improve the agreement with the diffraction data. In contrast, no crystallographic defects or site disorder were detected within the sensitivity of the present structural refinements. These observations indicate that the local symmetry breaking in Phase I is an intrinsic property of NbSeI, rather than being induced by structural defects or disorder, as has been discussed for $LiNiO_2$[20-23] and $Ba_3CuSb_2O_9$[54-57].

Notably, the diffuse scattering observed in Phase I is substantially weaker than that typically found in pyrochlore-related systems exhibiting pronounced short-range correlations along the ⟨110⟩ directions, i.e., along the edges of the tetrahedra[58]. Thus, the present diffraction data do not provide evidence for a conventional, strongly correlated ⟨110⟩-type distortion pattern. Instead, they are consistent with local Jahn–Teller-like symmetry breaking of the $Nb_4$ tetrahedra without cooperative long-range order. The absence of cooperative long-range order may originate from intrinsic effects, such as geometrical frustration or competing interactions[45,59], although its microscopic origin remains to be clarified. Furthermore, the present experimental evidence for local symmetry breaking does not distinguish between static and dynamic local distortions. Dynamic probes will therefore be required to determine the time-dependent character of the local symmetry breaking.

More generally, electronically degenerate systems can lower their symmetry through lattice distortions, bond disproportionation, magnetic ordering, or defect formation, thereby avoiding an unstable metallic state.[60,61] The significance of NbSeI therefore lies not in symmetry lowering itself, but in revealing two distinct structural responses associated with molecular-orbital degeneracy lifting in a tetrahedral $Nb_4$ cluster system. These responses are associated with the suppression of the flat-band metallic state predicted for the cubic average structure and the emergence of a nonmagnetic insulating state.

## Conclusion

NbSeI exhibits nonmagnetic insulating behavior over the entire measured temperature range up to 500 K, associated with two distinct structural responses accompanying the lifting of molecular-orbital degeneracy in the $Nb_4$ tetrahedral clusters. Phase II below $T_s$ = 106 K is a molecular-orbital-ordered insulator in which all the $Nb_4$ tetrahedra distort into a roughly 3-in-1-out configuration. In Phase I above $T_s$, the average crystal structure retains cubic symmetry, but the molecular-orbital degeneracy is most likely lifted by local symmetry breaking associated with pronounced local distortions of the $Nb_4$ tetrahedra without cooperative long-range order, thereby suppressing the average-structure flat-band metal predicted by the first-principles calculations. Future studies aimed at suppressing these insulating states through external fields, pressure, or chemical substitution may provide access to further electronic phases, potentially including the flat-band metal as well as other quantum phases arising from the molecular-orbital degrees of freedom.

## Experimental section

### Sample synthesis

Single crystalline samples of NbSeI were synthesized by the following method. 1 : 1 : 1.05 molar ratio of Nb (99.9%, Rare Metallic), Se (99.999%, Kojundo Chemical Laboratory), and I (99.8%, Wako Pure Chemical Corporation) were sealed in evacuated quartz tubes. The tubes were heated and maintained at 1323 K for 72 h, slowly cooled to 1073 K at a rate of 2 K $h^{-1}$, and then furnace cooled to room temperature. The obtained samples were immersed and rinsed in ethanol until the ethanol was no longer colored.

### Physical property measurements

Magnetization measurements were performed using a MPMS-3 (Quantum Design). Electrical resistivity was measured by the four-point probe method. Heat capacity was measured by the relaxation method using a Physical Property Measurement System (Quantum Design). Dielectric constant on the single crystal sample was measured by a conventional two-terminal method using an LCR meter (E4980A, Agilent).

### x-ray diffraction measurement

Single crystal XRD experiments were performed at BL02B1 in SPring-8 in the temperature range of 50–300 K using 40 keV X-rays. The sample temperature was controlled by a $N_2$ and He gas blowing system. Diffraction data were corrected using 2D semiconductor detector PILATUS 3X CdTe (DECTRIS Ltd, Switzerland). Indexing and extraction of the intensity were performed using CrysAlis Pro[62]. Averaging of diffraction intensity data, initial structure construction, and structure refinement were performed using JANA2006[63]. VESTA[64] was used to draw the crystal structures. The cif files of Phase I and II have been uploaded to the CCDC database with the deposition number of 2504335 and 2504334, respectively.

Powder X-ray diffraction data for PDF analysis at 300 K were collected at the BL04B2 beamline of SPring-8. To minimize the background arising from fluorescence X-rays, an X-ray wavelength of 0.11 Å and Ge point detectors were used. The obtained diffraction data were converted into PDF data using software developed for the BL04B2 beamline[65]. The Fourier transformation was performed using the following parameters, selected by considering the trade-off between statistical accuracy and real-space resolution: $Q_{\mathrm{min}} = 0.3$ Å$^{-1}$, $Q_{\mathrm{MAX}} = 25.0$ Å$^{-1}$, and $\delta Q = 0.01$ Å$^{-1}$. Structural fitting of the PDF data was performed using PDFgui[66].

### Theoretical calculations

First-principles calculations were performed based on density functional theory (DFT) using the full-potential linearized augmented plane wave (FP-LAPW) method implemented in the WIEN2k code[67]. The exchange-correlation effects were treated within the generalized gradient approximation (GGA) using the Perdew–Burke–Ernzerhof (PBE) functional. The total energy convergence criterion was set to 0.0001 Ry. For Phase I, the calculations were performed for the primitive unit cell (one fourth of the conventional cell shown in Fig. 1(a)) with a 20 × 20 × 20 $k$-point mesh in the Brillouin zone. For Phase II, a 12 × 12 × 6 $k$-point mesh was used. The experimental crystal structures measured at 300 K for Phase I and 100 K for Phase II were adopted in the calculations.

Maximally localized Wannier functions (MLWFs) were constructed to obtain a tight-binding model representing the DFT band structure of NbSeI. The Bloch states near the Fermi level were projected onto Nb $a_1$ orbitals to build the Wannier basis using the WANNIER90 code[68] interfaced with WIEN2k. After convergence, the obtained hopping parameters were $t_{\mathrm{n}} = -0.42$ eV for the nearest-neighbor and $t_{nn} = -0.05$ eV for the next-nearest-neighbor hoppings, while the magnitudes of all other hoppings were less than 0.04 eV. To simulate the distortion of Nb tetrahedra, only $t_n$ for each bond was scaled.

## Acknowledgments

The authors are grateful to K. Kuroda, Y. Fujisawa, K. Aoyama, M. Gen, Y. Kohama, Y. Ejiri, and J. Fujioka for the helpful discussions and N. Katayama and T. Hara for their help with the X-ray diffraction measurements. XRD experiments were conducted at the BL02B1 and BL04B2 of SPring-8, Hyogo, Japan (Proposals No. 2024B0304, No. 2024B1585, No. 2025A1631, No. 2025A1950, and 2025B1306).

## Funding

This study was supported by JSPS KAKENHI (Grant Nos.: 23H01831, 23K26524, 24H01644, 24K01329, 24K06938, 24K17006, 26H00590, 26H00640, 26H02231, and 26K17087) and JST ASPIRE (Grant No. JPMJAP2314).

## Competing interests

The authors declare no competing interests.

## Associated Content

**Supporting Information (PDF)**

I. Results of single crystal XRD analysis and structural parameters of Phase I (300 K)

II. Results of single-crystal XRD analysis and structural parameters for Phase II (100 K)

III. The electronic structure using the tight-binding model with twenty and four orbitals

IV. Phonon calculations for the cubic average structure

**Accession Codes**

Crystallographic data for Phases I and II have been deposited in the Cambridge Crystallographic Data Centre (CCDC) under deposition numbers 2504335 and 2504334.

## Data and materials availability

All data needed to evaluate the conclusions in the paper are present in the main text and/or the Supporting Information. Additional data related to this paper may be requested from the corresponding authors.

# TOC graphic

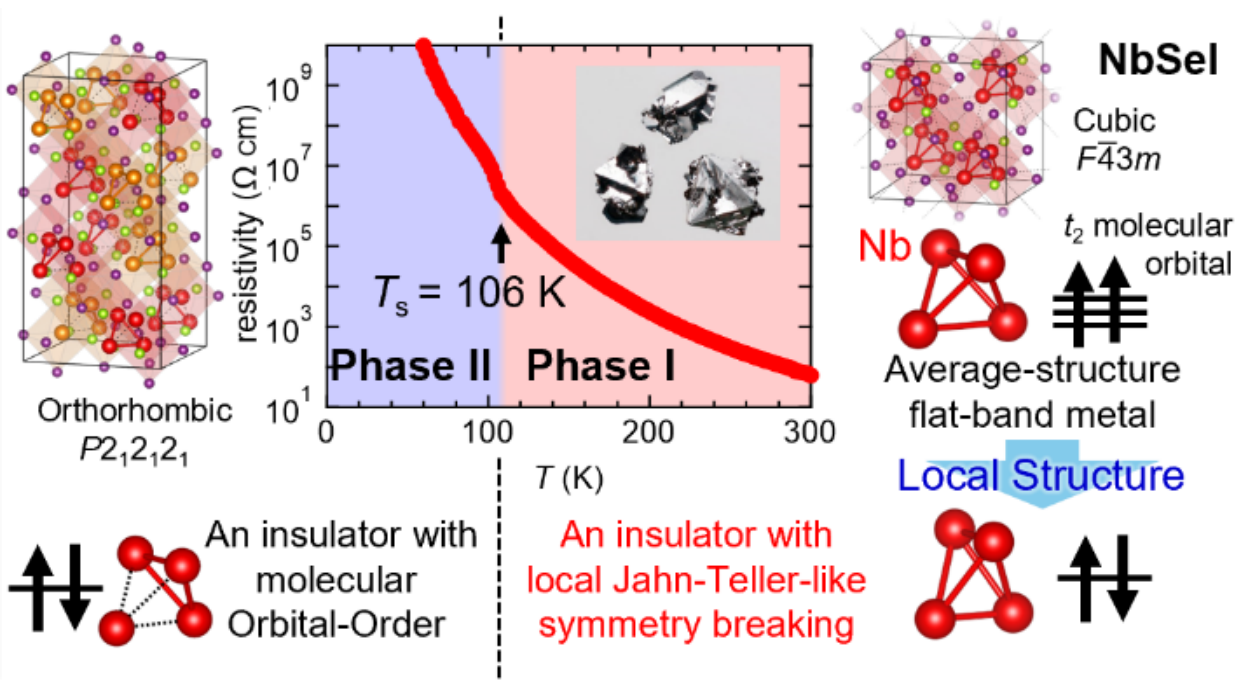

NbSeI
Cubic
$F\bar{4}3m$
$t_2$ molecular orbital
Nb
Average-structure flat-band metal
Local Structure
resistivity (Ω cm)
$T_s$ = 106 K
Phase II
Phase I
T (K)
Orthorhombic
$P2_12_12_1$
An insulator with molecular Orbital-Order
An insulator with local Jahn-Teller-like symmetry breaking

# Supporting Information of :
# "Molecular-Orbital Degeneracy Lifting and Suppression of a Flat-Band Metal in a Tetrahedral Cluster System NbSeI"

Keita Kojima[1,2]*, Youichi Yamakawa[3], Ryutaro Okuma[1], Shunsuke Kitou[4], Taishun Manjo[5], Hayato Takano[1], Jun-ichi Yamaura[1], Yusuke Tokunaga[4], Taka-hisa Arima[4], and Yoshihiko Okamoto[1]*

[1] Institute for Solid State Physics, the University of Tokyo, Kashiwa 277-8581, Japan

[2] Graduate School of Environment, Life, Natural Science and Technology, Okayama University, Okayama 700-8530, Japan

[3] Department of Physics, Nagoya University, Nagoya 464-8602, Japan

[4] Department of Advanced Materials Science, the University of Tokyo, Kashiwa 277-8561, Japan

[5] Japan Synchrotron Radiation Research Institute (JASRI), SPring-8, Hyogo 679-5198, Japan

* Corresponding author: keita-kojima@okayama-u.ac.jp
yokamoto@issp.u-tokyo.ac.jp

## I. Results of single crystal XRD analysis and structural parameters of Phase I (300 K)

The experimental conditions during the single-crystal XRD measurements of NbSeI at 300 K (Phase I), and the crystallographic parameters obtained by the structural analysis at 300 K are shown in Supplementary Tables 1 and 2, respectively. When the occupancy for each site is considered as a variable, they all converged to 1 within the experimental error, irrespective of the initial conditions, indicating the absence of vacancies and intersite defect at each site. Therefore, the crystallographic parameters obtained when the occupancy of each site was fixed to be one are listed in Table 2.

**Supplementary Table 1**: Summary of crystallographic data of NbSeI at 300 K (Phase I).

| **$\lambda$ (Å)** | **$d_{min}$ (Å)** | **Laue class** | **$R_{merge}$ (3$\sigma$ cut)** | **Unique reflections (3$\sigma$ cut)** | **Total reflections (3$\sigma$ cut)** | **$R_{merge}$ (all)** | **Unique reflections (all)** | **Total reflections (all)** |
|---|---|---|---|---|---|---|---|---|
| 0.3094 | 0.28 | $m\bar{3}m$ | 3.84% | 2558 | 53120 | 3.84% | 2831 | 64679 |
| **Space Group** | **Complete-ness** | **Average Redundancy** | **$R$ (3$\sigma$ cut, $d_{max}$=1Å, #2467)** | **$wR$ (3$\sigma$ cut, $d_{max}$=1Å)** | **GoF (3$\sigma$ cut, $d_{max}$=1Å)** | **$R$ (all, $d_{max}$=1Å, #2740)** | **$wR$ (all, $d_{max}$=1Å)** | **GoF (all, $d_{max}$=1Å)** |
| $F\bar{4}3m$ | 99.14% | 22.847 | 3.29% | 6.27% | 4.46 | 3.57% | 6.29% | 4.26 |

***About Twins***

Twin matrix: $\begin{pmatrix} -1 & 0 & 0 \\ 0 & -1 & 0 \\ 0 & 0 & -1 \end{pmatrix}$ (inversion)

Volume fraction: 0.538(13):0.462

**Supplementary Table 2**: Structural parameters of NbSeI at 300 K (Phase I). The cif file has been uploaded to the CCDC database with the deposition number of 2504335. Occupancy for each site is fixed to be one. This is because when each parameter was considered as a variable, they all converged to 1 within the range of error.

| **Space group** | **Z** | **Lattice parameters** | | | | | |
|---|---|---|---|---|---|---|---|
| | | ***a*** **(Å)** | ***b*** **(Å)** | ***c*** **(Å)** | ***α*** **(°)** | ***β*** **(°)** | ***γ*** **(°)** |
| $F\bar{4}3m$ | 16 | 10.7788(1) | 10.7788(1) | 10.7788(1) | 90 | 90 | 90 |
| **Atomic coordination** | **Element** | **Site** | ***x*** | ***y*** | ***z*** | $\boldsymbol{U_{eq.}}$ **(Å$^{-2}$)** | **Occ.** |
| | Nb | 16*d* | 0.347597(15) | = *x* | = *x* | 0.01382(4) | 1 |
| | Se | 16*d* | 0.11742(3) | = *x* | = *x* | 0.01485(4) | 1 |
| | I | 16*d* | 0.623699(8) | = *x* | = *x* | 0.01150(3) | 1 |
| **Anisotropic Temperature factor** | **Element** | $\boldsymbol{U_{11}}$ | $\boldsymbol{U_{22}}$ | $\boldsymbol{U_{33}}$ | $\boldsymbol{U_{12}}$ | $\boldsymbol{U_{13}}$ | $\boldsymbol{U_{23}}$ |
| | Nb | 0.01382(4) | = $U_{11}$ | = $U_{11}$ | 0.00074(3) | = $U_{12}$ | = $U_{12}$ |
| | Se | 0.01485(4) | = $U_{11}$ | = $U_{11}$ | 0.00078(4) | = $U_{12}$ | = $U_{12}$ |
| | I | 0.01150(3) | = $U_{11}$ | = $U_{11}$ | -0.000429(13) | = $U_{12}$ | = $U_{12}$ |

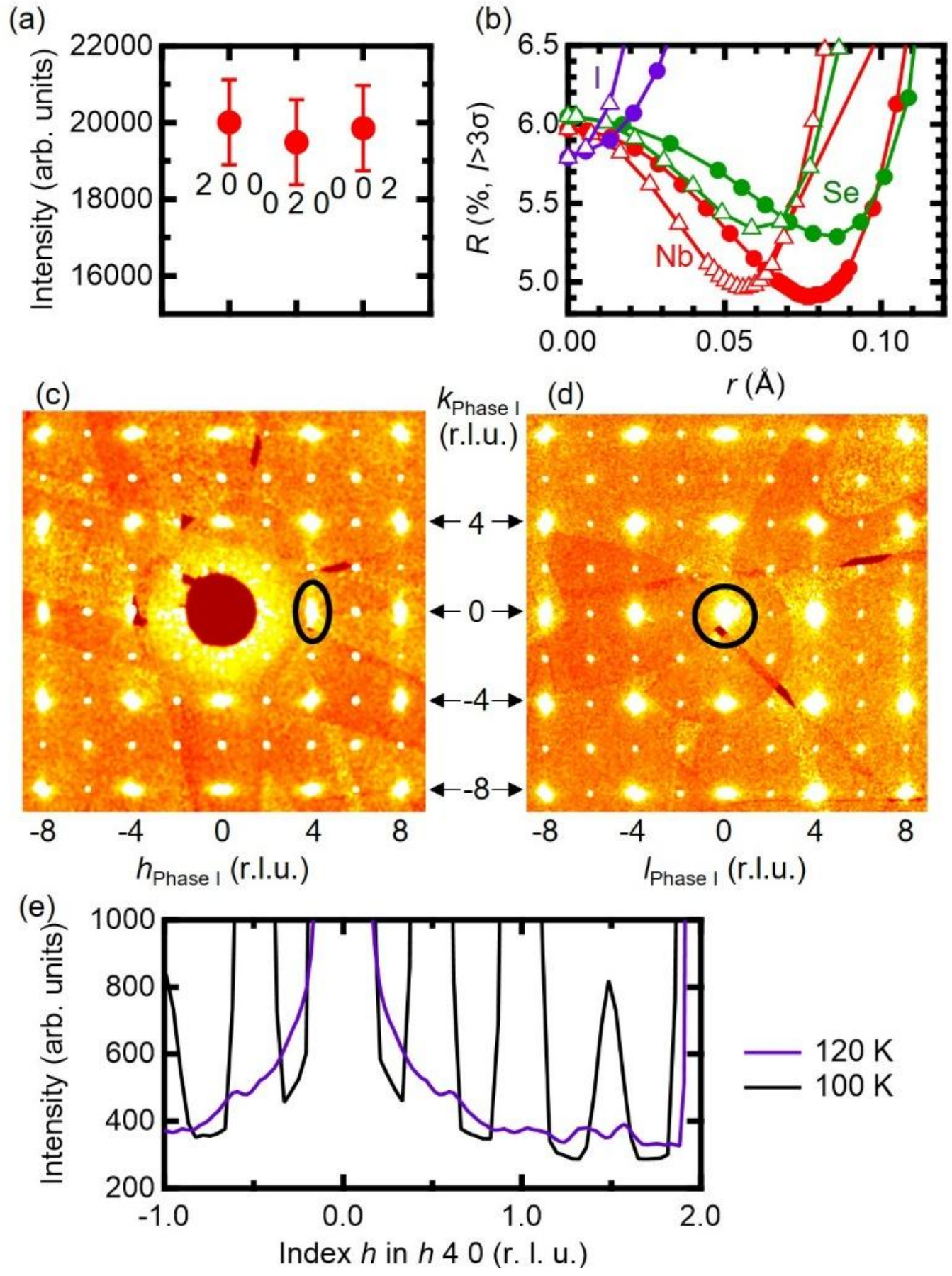

**Supplementary Fig. 1:** (a) Intensities of 200, 020 and 002 in single-crystal XRD of Phase I (300 K). (b) The reliability factor in the structural refinement according to the split site model for the 160 K data of Nb, Se, and I sites, as a function of displacement from the regular position. Distortions along the <111> and <110> directions are indicated by open triangles and filled circles, respectively. (c,d) Single-crystal XRD patterns of (c) *hk*0 plane and (d)

4*kl* plane taken at 300 K. 4 0 0 Bragg peak is highlighted by a black ellipse or circle. (e) XRD line profiles of 120 K and 100 K data at *h* 4 0.

Supplementary Figure 1 (a) shows the intensities of 200, 020, and 002 reflections measured at 300 K. Intensities of these reflections coincide within the experimental error, supporting that Phase I of NbSeI has an average cubic symmetry.

Supplementary Figure 1 (b) shows the results of the split-site model analysis applied to Se and I sites. Data for the Nb site, shown in Fig. 4 (c) of the main article, are also included for reference. As shown in Supplementary Table 2, little anisotropy is observed in the atomic displacement parameters. However, this is unlikely to reflect the intrinsic nature of Phase I. Since the average cubic symmetry is maintained in Phase I, local atomic displacements that break the cubic symmetry cannot be represented by the crystallographic parameters, and such anisotropic displacements may be spatially averaged and therefore appear isotropic. Therefore, we performed a split-site model analysis considering atomic displacements along the ⟨110⟩ directions, which locally break cubic symmetry, as well as displacements along the ⟨111⟩ directions preserving the cubic symmetry. In the ⟨110⟩ case, displacements not only along the [110] but also the [101] and [011] directions were assumed to retain cubic symmetry as a whole. For the Se case, the *R* value decreases by increasing the displacement from zero and reaches the minimum value at similar displacement values as those for Nb. In contrast, the *R* value monotonically increases with increasing the displacement for the I case. These behaviors suggest that Nb and Se atoms are simultaneously displaced in a $Nb_4Se_4$ cube.

Supplementary Figures 1 (c) and (d) show single-crystal XRD patterns of the *hk*0 and 4*kl* planes, respectively, taken at 300 K. The diffuse scattering appears around the strong Bragg peaks, which may evidence short-range correlations of the local distortions in Phase I. However, the intensity of the diffuse scattering is weak and it is not observed around weak Bragg peaks. The observed diffuse scattering may also originate from thermal excitation of phonons. These results suggest that a definitive discussion on the origin of diffuse scattering based solely on the present data is impossible.

It is also noted that the line profiles along *h* direction around 0 4 0 for 120 K and 100 K shown in Supplementary Figure 1 (e) reveal a reduction of the background intensity around the Bragg peak at 100 K, corresponding to Phase II. This observation suggests that the diffuse scattering most likely appears only in Phase I.

## II. Results of single-crystal XRD analysis and structural parameters for Phase II (100 K)

The experimental conditions during the single-crystal XRD measurements of NbSeI at 100 K (Phase II), and the crystallographic parameters obtained by the structural analysis at 100 K are shown in Supplementary Tables 3 and 4, respectively. Table 5 provides the twin transformation matrices and the volume fraction of each twin. Since peak splitting was not observed in the single-crystal XRD measurements, the lattice parameters are shown under the condition of $a = b = c/2$. Given the large number of twins and the relatively weak intensity of the superlattice reflections, we assumed the same isotropic temperature factor for each atom in this analysis.

**Supplementary Table 3**: Summary of crystallographic data of NbSeI at 100 K (Phase II). When the intensities are extracted in CrysAlisPro with multiple domains, the total reflection information appears to be lost upon import into Jana2006, and all reflections are merged. As a result, $R_{merge}$ and average redundancy cannot be defined.

| **$\lambda$ (Å)** | **$d_{min}$ (Å)** | **Laue class** | **$R_{merge}$ (3$\sigma$ cut)** | **Unique reflections (3$\sigma$ cut)** | **Total reflections (3$\sigma$ cut)** | **$R_{merge}$ (all)** | **Unique reflections (all)** | **Total reflections (all)** |
|---|---|---|---|---|---|---|---|---|
| 0.30931 | 0.28 | 222 | N/A | 113227 | N/A | N/A | 224862 | N/A |
| **Space Group** | **Complete-ness** | **Average Redundancy** | **$R$ (3$\sigma$ cut)** | **$wR$ (3$\sigma$ cut)** | **GoF (3$\sigma$ cut)** | **$R$ (all)** | **$wR$ (all)** | **GoF (all)** |
| $P2_12_12_1$ | 97.33% | N/A | 4.88% | 4.93% | 2.07 | 6.82% | 5.02% | 1.50 |

**Supplementary Table 4**: Structural parameters of NbSeI at 100 K (Phase II). The cif file has been uploaded to the CCDC database with the deposition number of 2504334. The occupancy at each atomic site is fixed to be 1.

| **Space group** | **Z** | **Lattice parameters** | | | | | |
|---|---|---|---|---|---|---|---|
| | | ***a* (Å)** | ***b* (Å)** | ***c* (Å)** | ***α* (°)** | ***β* (°)** | ***γ* (°)** |
| $P2_12_12_1$ | 32 | 10.7413(4) | 10.7413(4) | 21.4826(7) | 90 | 90 | 90 |
| **Atomic coordination** | **Element** | **Site** | ***x*** | ***y*** | ***z*** | **$U_{eq.}$ (Å$^{-2}$)** | **Occ.** |
| | Nb 1 | 4*a* | 0.600721(18) | 0.344369(19) | 0.670497(11) | 0.005028(5) | 1 |
| | Nb 2 | 4*a* | 0.902907(17) | 0.33628(2) | 0.826676(9) | = Nb 1 | 1 |
| | Nb 3 | 4*a* | 0.589620(18) | 0.35601(2) | 0.175762(10) | = Nb 1 | 1 |
| | Nb 4 | 4*a* | 0.893771(18) | 0.354879(17) | 0.323061(11) | = Nb 1 | 1 |
| | Nb 5 | 4*a* | 0.41053(2) | 0.34351(2) | 0.578204(8) | = Nb 1 | 1 |
| | Nb 6 | 4*a* | 0.096472(19) | 0.35280(2) | 0.927916(8) | = Nb 1 | 1 |
| | Nb 7 | 4*a* | 0.39817(2) | 0.35177(2) | 0.079674(8) | = Nb 1 | 1 |
| | Nb 8 | 4*a* | 0.10512(2) | 0.34294(2) | 0.427586(8) | = Nb 1 | 1 |
| | Se 1 | 4*a* | 0.87080(3) | 0.62248(3) | 0.558761(16) | 0.005382(6) | 1 |
| | Se 2 | 4*a* | 0.62712(3) | 0.61070(3) | 0.938366(14) | = Se 1 | 1 |
| | Se 3 | 4*a* | 0.85873(3) | 0.61253(3) | 0.055768(14) | = Se 1 | 1 |
| | Se 4 | 4*a* | 0.63423(3) | 0.61993(3) | 0.442205(14) | = Se 1 | 1 |
| | Se 5 | 4*a* | 0.13929(3) | 0.60641(3) | 0.695108(11) | = Se 1 | 1 |
| | Se 6 | 4*a* | 0.37434(3) | 0.61158(3) | 0.811867(11) | = Se 1 | 1 |
| | Se 7 | 4*a* | 0.13094(3) | 0.62431(3) | 0.193819(11) | = Se 1 | 1 |
| | Se 8 | 4*a* | 0.36577(2) | 0.62498(3) | 0.312597(10) | = Se 1 | 1 |
| | I 1 | 4*a* | 0.871057(14) | 0.630368(14) | 0.309836(8) | 0.003996(3) | 1 |
| | I 2 | 4*a* | 0.622943(15) | 0.630474(14) | 0.187913(8) | = I 1 | 1 |
| | I 3 | 4*a* | 0.877643(15) | 0.615883(15) | 0.811842(8) | = I 1 | 1 |
| | I 4 | 4*a* | 0.631719(15) | 0.618688(15) | 0.686681(8) | = I 1 | 1 |
| | I 5 | 4*a* | 0.129226(18) | 0.621425(16) | 0.440906(6) | = I 1 | 1 |
| | I 6 | 4*a* | 0.368641(17) | 0.627172(16) | 0.063978(6) | = I 1 | 1 |
| | I 7 | 4*a* | 0.122097(19) | 0.626329(16) | 0.940146(6) | = I 1 | 1 |
| | I 8 | 4*a* | 0.377165(15) | 0.620112(17) | 0.564574(6) | = I 1 | 1 |

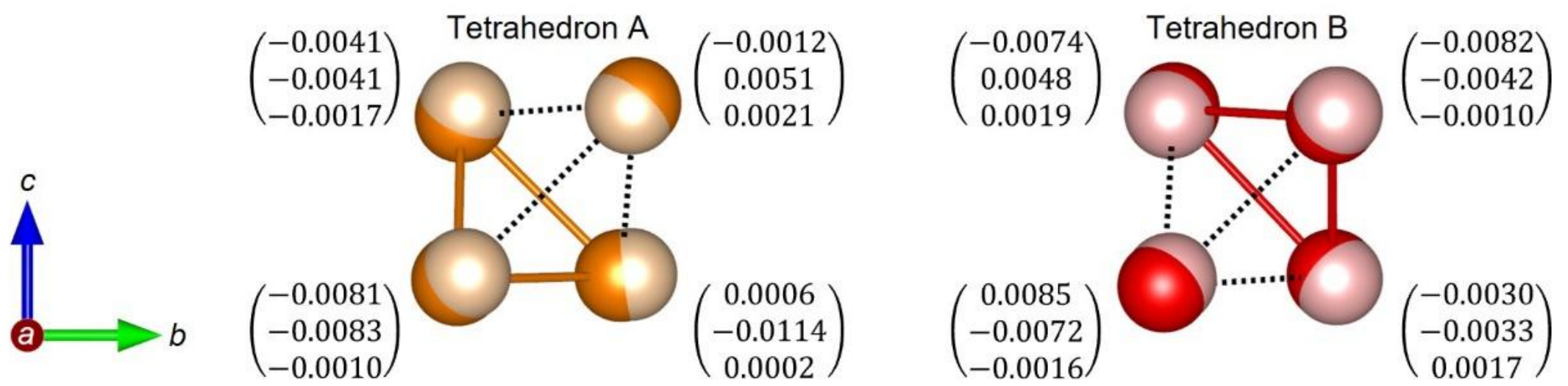

**Supplementary Fig. 2:** Nb atomic positions in Phase II (deep) compared to those in Phase I (light). Vectors in the figure represent the changes in atomic coordination from Phase I.

**Supplementary Table 5**: Twin matrixes and the volume fraction at 100 K (phase II)

| Twin No. | Matrix | Vol. fraction | Operation detail |
|---|---|---|---|
| 1 | $\begin{pmatrix} 1 & 0 & 0 \\ 0 & 1 & 0 \\ 0 & 0 & 1 \end{pmatrix}$ | 0.075(5) | Identity operation |
| 2 | $\begin{pmatrix} 1 & 0 & 0 \\ 0 & 0 & -0.5 \\ 0 & 2 & 0 \end{pmatrix}$ | 0.0859(13) | Four-fold rotation axis in direction $[\bar{1}\ 0\ 0]$ |
| 3 | $\begin{pmatrix} 0 & 0 & 0.5 \\ 0 & 1 & 0 \\ -2 & 0 & 0 \end{pmatrix}$ | 0.0645(13) | Four-fold rotation axis in direction $[0\ \bar{1}\ 0]$ |
| 4 | $\begin{pmatrix} 0 & 1 & 0 \\ -1 & 0 & 0 \\ 0 & 0 & 1 \end{pmatrix}$ | 0.1357(15) | Four-fold rotation axis in direction $[0\ 0\ 1]$ |
| 5 | $\begin{pmatrix} 0 & 1 & 0 \\ 0 & 0 & -0.5 \\ -2 & 0 & 0 \end{pmatrix}$ | 0.0986(13) | Three-fold rotation axis direction $[\bar{2}\ \bar{2}\ 1]$<br>(Four-fold rotation axis in direction $[0\ 0\ 1]$ for twin No. 2) |
| 6 | $\begin{pmatrix} 0 & 0 & 0.5 \\ -1 & 0 & 0 \\ 0 & -2 & 0 \end{pmatrix}$ | 0.0774(13) | Three-fold rotation axis direction $[2\ \bar{2}\ 1]$<br>(Four-fold rotation axis in direction $[0\ 0\ 1]$ for twin No. 3) |
| 7 | $\begin{pmatrix} -1 & 0 & 0 \\ 0 & -1 & 0 \\ 0 & 0 & -1 \end{pmatrix}$ | 0.097(3) | Inversion operation |
| 8 | $\begin{pmatrix} -1 & 0 & 0 \\ 0 & 0 & 0.5 \\ 0 & -2 & 0 \end{pmatrix}$ | 0.0794(13) | Inverse four-fold rotation axis in direction $[\bar{1}\ 0\ 0]$<br>(Inversion operation for twin No. 2) |
| 9 | $\begin{pmatrix} 0 & 0 & -0.5 \\ 0 & -1 & 0 \\ 2 & 0 & 0 \end{pmatrix}$ | 0.0663(13) | Inverse four-fold rotation axis in direction $[0\ \bar{1}\ 0]$<br>(Inversion operation for twin No. 3) |
| 10 | $\begin{pmatrix} 0 & -1 & 0 \\ 1 & 0 & 0 \\ 0 & 0 & -1 \end{pmatrix}$ | 0.0668(15) | Inverse four-fold rotation axis in direction $[0\ 0\ 1]$ |
| 11 | $\begin{pmatrix} 0 & -1 & 0 \\ 0 & 0 & 0.5 \\ 2 & 0 & 0 \end{pmatrix}$ | 0.0901(13) | Inverse three-fold rotation axis direction $[\bar{2}\ \bar{2}\ 1]$<br>(Inverse four-fold rotation axis in direction $[0\ 0\ 1]$ for twin No. 2) |
| 12 | $\begin{pmatrix} 0 & 0 & -0.5 \\ 1 & 0 & 0 \\ 0 & 2 & 0 \end{pmatrix}$ | 0.0631(13) | Inverse three-fold rotation axis direction $[2\ \bar{2}\ 1]$<br>(Inverse four-fold rotation axis in direction $[0\ 0\ 1]$ for twin No. 3) |

## III. The electronic structure using the tight-binding model with twenty and four orbitals

Supplementary Figure 3 shows the electronic structures of NbSeI calculated using the tight-binding model based on first-principles calculation with Wannier90 code[57]. Supplementary Figure 3 (a) shows the result of the twenty-orbital case, which considers all 4*d* orbitals of four Nb atoms in a tetrahedron. This tight-binding result well reproduces the first-principles calculation results between –2 eV and 4 eV, indicating that the Nb 4*d* orbitals are dominant in the electronic states in this energy region. The flat dispersion at around $E_F$ is also well reproduced. In the tight-binding results, there are single, double, and triple dispersions at around –2 eV, –1 eV, and $E_F$, corresponding to the single, doubly-degenerated, and triply-degenerated molecular orbitals of a $Nb_4$ tetrahedron, respectively, as shown in Supplementary Figure 3 **a**. The electronic states between 1 eV and 4 eV correspond to the two triply-degenerated molecular orbitals with higher energies, in addition to $e_g$ atomic orbitals. Moreover, as shown in Supplementary Figure 3 (b) [same as Figure 5 (a) in the main article], single and triple dispersions at –2 eV and $E_F$, respectively, are well reproduced in the tight-binding result using four $a_1$ atomic orbitals in a $Nb_4$ tetrahedron. This result reflects that the single and triple dispersions correspond to the bonding and antibonding molecular orbitals made of $a_1$ atomic orbitals, as shown in Supplementary Figure 3 (b).

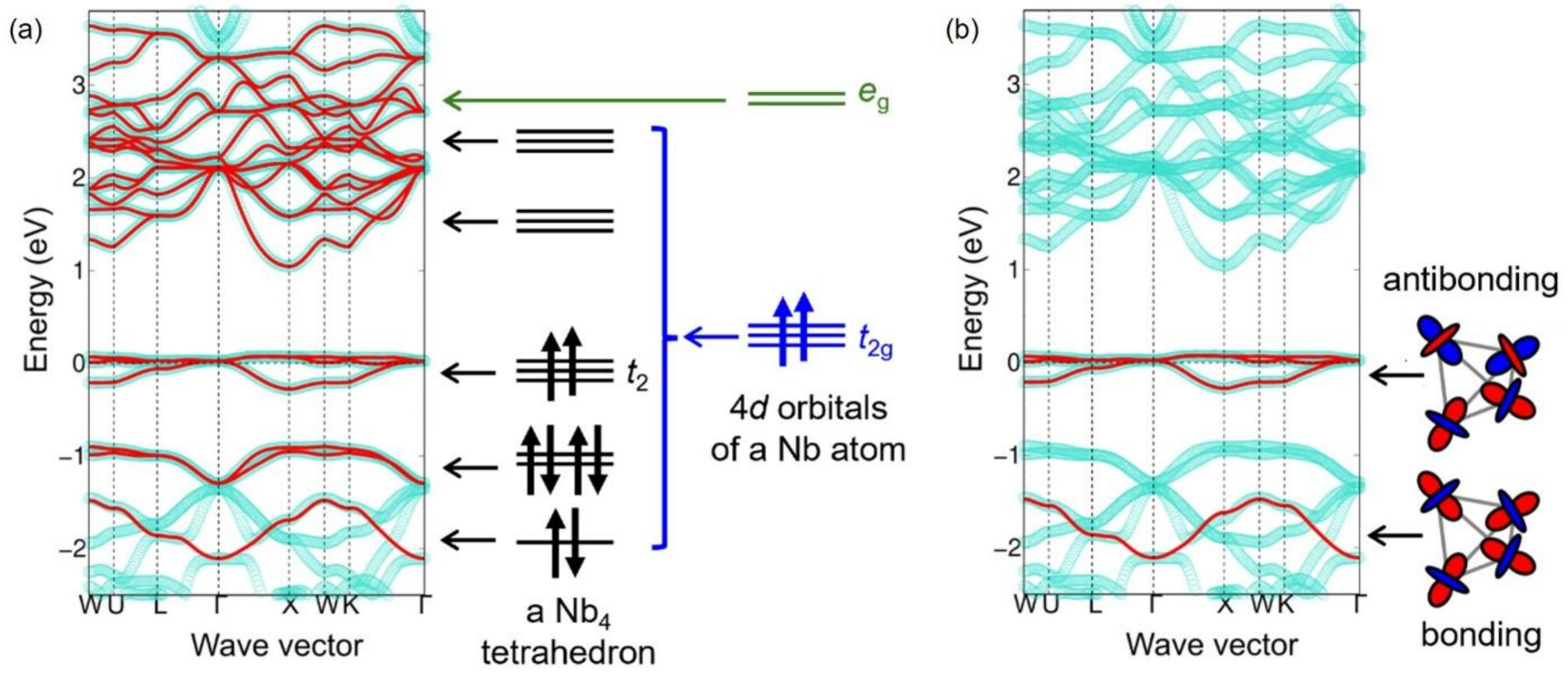

**Supplementary Fig. 3**: The electronic structure based on the tight-binding model using (a) twenty orbitals, which are all Nb 4*d* orbitals in a $Nb_4$ tetrahedron, and (b) four orbitals, which consist of $a_1$ atomic orbitals of each Nb atom of a $Nb_4$ tetrahedron. The red and cyan curves are the tight-binding and the first principles calculation results, respectively. The first principles calculation data are calculated without spin–orbit coupling using refined structural parameters at 300 K.

## IV. Phonon calculations for the cubic average structure

Harmonic phonon calculations were performed for the cubic average structure of Phase I using density functional theory as implemented in the VASP code[S1] in combination with the Phonopy package[S2]. The exchange-correlation functional was described within the generalized gradient approximation in the Perdew–Burke–Ernzerhof (PBE) form[S3]. A 4×4×4 Monkhorst-Pack *k*-point mesh[S4] and a plane-wave cutoff energy of 650 eV were employed for structural relaxation, ensuring convergence with respect to these parameters. Phonon calculations were performed using the finite-displacement method with a 2×2×2 supercell.

The calculated phonon dispersion is shown in Supplementary Fig. 4. Three phonon branches exhibit imaginary frequencies over broad regions of the Brillouin zone. No isolated minimum is observed at a particular wave vector, including the X point corresponding to the superlattice wave vector of Phase II. Therefore, within the present harmonic calculation, a unique unstable wave vector or displacement eigenvector cannot be identified.

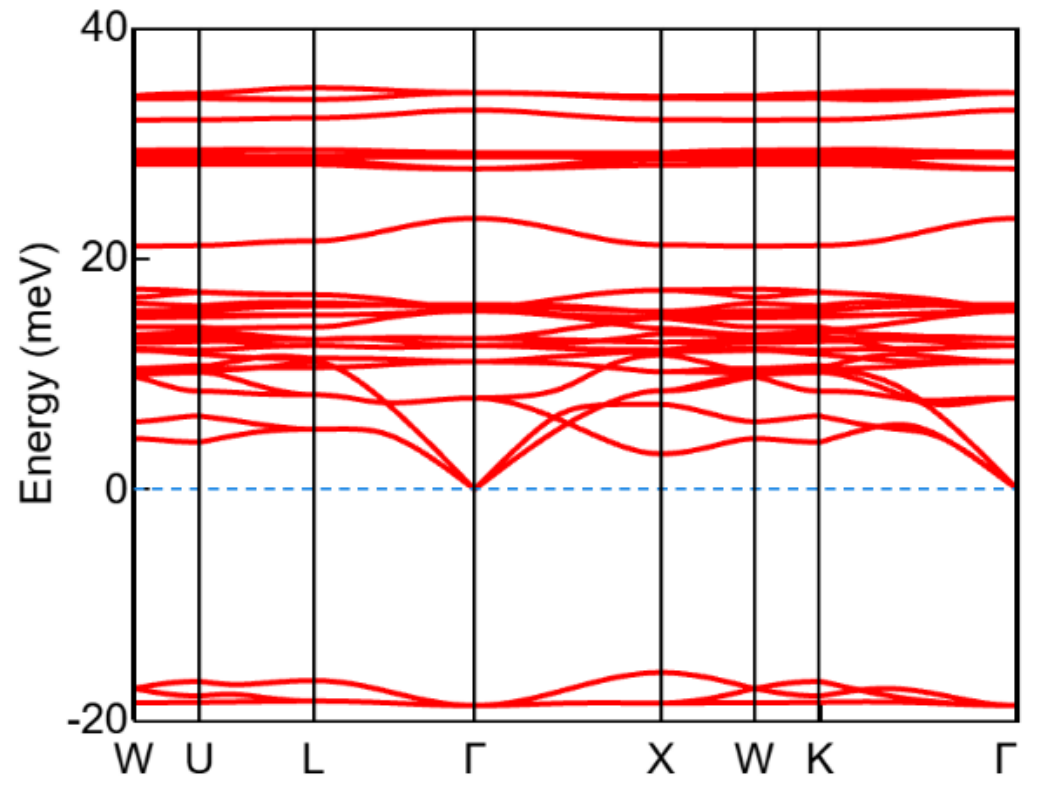

**Supplementary Fig. 4.** Calculated phonon dispersion of the cubic average structure of NbSeI. Imaginary frequencies are plotted as negative values.